\documentclass[final]{svjour3}
\usepackage{graphicx}
\usepackage{rotating}
\usepackage{amssymb}
\usepackage{mathptmx}
\usepackage{deluxetable}
\usepackage{comment}
\usepackage{hyphenat}

\makeatletter
\journalname{Journal of Low Temperature Physics}

\begin{document}

\newcommand{\hdblarrow}{H\makebox[0.9ex][l]{$\downdownarrows$}-}
\title{The 90 and 150~GHz universal focal-plane modules for the Simons Observatory}

\author{Heather McCarrick$^1$
    \and Kam Arnold$^2$
    \and Zachary Atkins$^1$
    \and Jason Austermann$^3$
    \and Tanay Bhandarkar$^4$
    \and Steve K. Choi$^{5,6}$
    \and Cody J. Duell$^5$
    \and Shannon M. Duff$^3$
    \and Daniel Dutcher$^1$
    \and Nicholas Galitzki$^2$
    \and Erin Healy$^1$
    \and Zachary B. Huber$^5$
    \and Johannes Hubmayr$^3$
    \and Bradley R. Johnson$^7$
    \and Michael D. Niemack$^{5,6,8}$
    \and Joseph Seibert$^{2}$
    \and Maximiliano Silva-Feaver$^2$
    \and Rita F. Sonka$^1$
    \and Suzanne T. Staggs$^1$
    \and Eve M. Vavagiakis$^{5}$
    \and Yuhan Wang$^1$
    \and Zhilei Xu$^{9}$
    \and Kaiwen Zheng$^1$
    \and Ningfeng Zhu$^4$}

\institute{1. Joseph Henry Laboratories of Physics, Jadwin Hall, Princeton University, Princeton, NJ 08544, USA \\
2. Department of Physics, University of California San Diego, La Jolla, CA 92093, USA \\
3. Quantum Sensors Group, NIST, 325 Broadway, Boulder, CO 80305, USA \\
4. Department of Physics and Astronomy, University of Pennsylvania, 209 S 33rd St. Philadelphia, PA 19104, USA\\
5. Department of Physics, Cornell University, Ithaca, NY 14853, USA\\
6. Department of Astronomy, Cornell University, Ithaca, NY 14853, USA\\
7. Department of Astronomy, University of Virginia, Charlottesville, VA 22904, USA \\
8.  Kavli Institute at Cornell for Nanoscale Science, Cornell University, Ithaca, NY 14853, USA\\
9. MIT Kavli Institute, Massachusetts Institute of Technology, 77 Massachusetts Avenue, Cambridge, MA 02139, USA\\
\email{hm8@princeton.edu}}

\maketitle

\begin{abstract}

The Simons Observatory (SO) is a suite of telescopes located in the Atacama Desert in Chile that will make sensitive measurements of the cosmic microwave background. There are a host of cosmological and astrophysical questions that SO is forecasted to address. The universal focal-plane modules (UFMs) populate the four SO telescope receiver focal planes. There are three varieties of UFMs, each of which contains transition-edge-sensor bolometers observing in two spectral bands between 30 and 290~GHz. We describe the novel mid-frequency UFMs, which target two of the six spectral bands at 90 and 150~GHz and are central to the cosmological goals of SO. 


\end{abstract}

\section{Introduction}

The Simons Observatory (SO) will observe the cosmic microwave background (CMB) with exceptional sensitivity from the Chilean Atacama Desert. The cosmological forecasts for SO have been laid out previously~\cite{so_forecast_2019}. SO will consist of four telescopes: one 6~m large-aperture~\cite{zhu_2021,xu2021} and three 0.5~m small-aperture telescopes~\cite{galitzki_2018}. The detectors will observe in six spectral bands in the millimeter-wave regime. 

\begin{figure}[t]
\begin{center}
\includegraphics[width=0.7\linewidth]{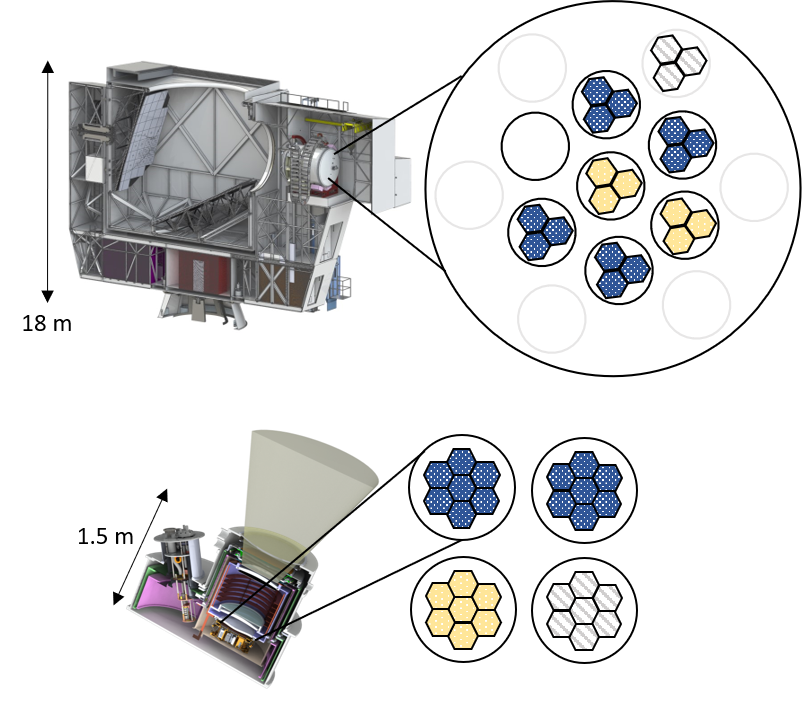}
\caption{Schematic showing the mid-frequency universal-focal plane (UFM) distribution among the SO receivers. {\it Top} The large-aperture telescope receiver (LATR) will host four mid-frequency optics tubes ({\it blue}), which will be complemented by two ultra-high frequency ({\it yellow}) and one low-frequency ({\it diagonal stripes}) optics tubes. {\it Bottom} Of the four small-aperture telescope (SAT) focal planes that will populate the three SATs, two will be mid-frequency.  (Color figure online.)}
\end{center}
\label{fig_so}
\end{figure}

Advancing experimental cosmology has required significant instrumentation development in order to achieve markedly improved sensitivity. The detector and readout research has been critical; these advances have made possible the planned deployment of the $>60,000$ transition-edge sensors (TES) bolometers. The detector array and cold readout multiplexer are packaged into a universal focal-plane module (UFM), which is a common assembly for the detector arrays across observed spectral frequencies and aforementioned telescope receivers. The detector arrays are dichroic, each observing in two frequencies, and are grouped in the following ways: low-frequency (LF) [30,40~GHz], mid-frequency (MF) [90, 150~GHz], and ultra-high-frequency (UHF) [220, 290~GHz]. The distribution of the arrays within SO is shown in Fig.~\ref{fig_so}. 

We have previously published the overall design and architecture of the UFM and demonstrated its full functionality, including achieving 1000x multiplexing of TES bolometers at the noise required for the SO forecasts~\cite{mccarrick_2021}. In this paper, we describe the specifications and implementations particular to the MF UFM. An accompanying paper~\cite{healy_2021} in this issue describes details of the UHF UFM.

\section{Methods}

\begin{figure}[t!]
\begin{center}
\includegraphics[width=1\linewidth]{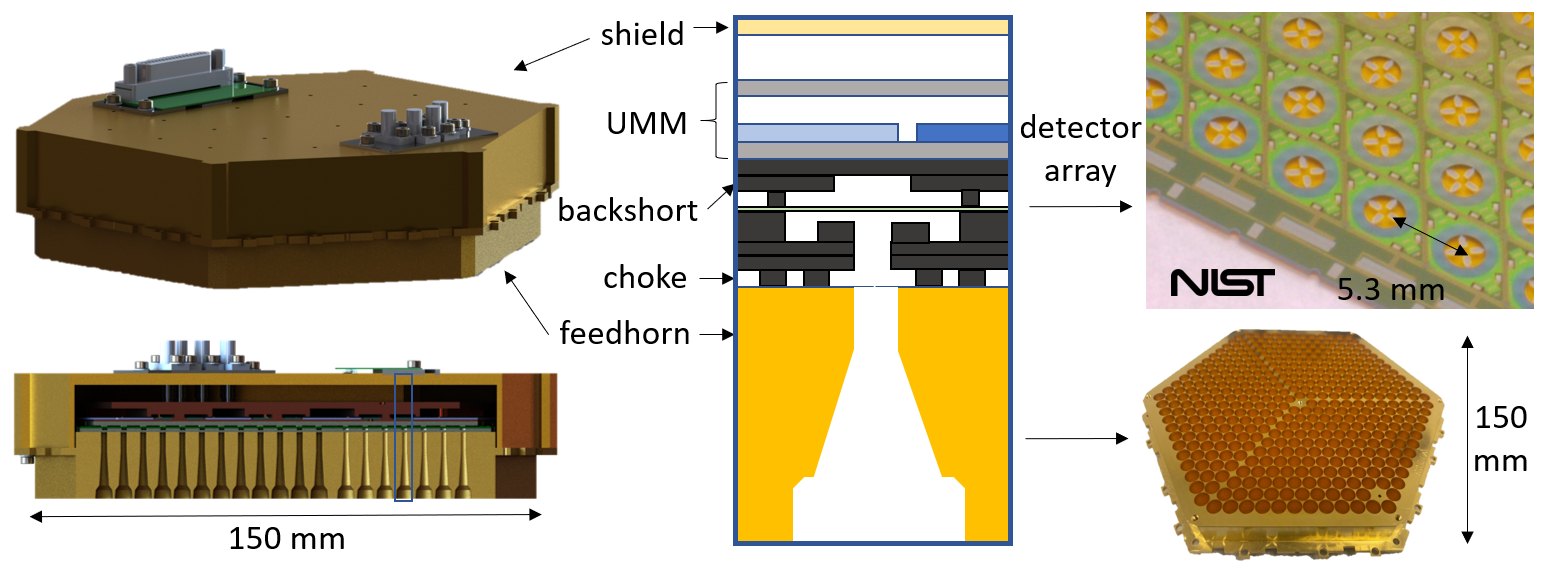}
\caption{{\it Left} Rendering of a mid-frequency UFM, with a cross-sectional view below. The gold-plated outer shield is visible, as are the RF and DC connectors that interface to the rest of the cryostat. Each MF UFM has 1728 optical TES bolometers. {\it Center} Cross-sectional view of a single element, corresponding to the outlined ({\it red}) component in the left figure. The spline-shaped feedhorn ({\it gold}) feeds a silicon detector stack ({\it grey}) consisting of the photonic choke, detector array, and backshort. Above the detector stack lies the cold multiplexer ({\it blue}), termed the universal $\mu$mux module (UMM)~\cite{mccarrick_2021}. An outer mechanical and magnetic shield ({\it yellow}) protects the inner pieces of the UFM. {\it Right} Photographs of a NIST MF TES bolometer array and an MF feedhorn array. Each detector rhombus corresponds to a single element. Within the dichroic pixel, the polarization-sensitive orthomode transducers each feed two 90 and 150~GHz TES bolometers.  (Color figure online.) }
\end{center}
\label{fig2}
\end{figure}

\subsection{The mid-frequency universal focal-plane module}
The common UFM architecture consists of a vertically stacked cold readout multiplexer, detector array, and optical coupling. These components are packaged into an approximately 150~mm diameter module, which can be tessellated to cover a focal plane. The MF UFM follows this prescription and is shown in the left panel of Fig.~\ref{fig2}. We describe the design below, starting from the sky-side. 

The MF optical coupling element is a feedhorn array made from gold-plated aluminum. Spline horns are chosen for their performance~\cite{simon_2018} and ease of fabrication with traditional direct machining methods, as opposed to the more labor-intensive micromachined platelet silicon feedhorns~\cite{Nibarger_2012}. Within the horn array, the feedhorn ends in a cylindrical waveguide. The optical path continues in the detector stack, discussed in the paragraph below. For completeness, the mechanical aspects of the horn array are important as well: the perimeter flange interfaces to the receiver and sets the alignment of the horn phase center, which is expected to be within 3~mm of the specified height as measured from the front of the receiver. Furthermore, the flange is designed to be interlocking with itself such that the space between tiled UFMs in minimized. The horns, like the outer shield behind the cold multiplexer, are gold-plated to increase the packaging thermal conductivity and facilitate cooling of the TES bolometers, which are held at 100~mK. 

In the middle of the UFM is the detector stack, which itself consists of a number of distinct components as shown in the center panel of Fig.~\ref{fig2}. The components are made of silicon, with all parts except the detector array gold-plated. The cylindrical waveguide in the horn array feeds a waveguide choke, followed by the detectors, and finally a quarter-wave backshort. The detector array has 1764 TES bolometers in a hex-packed configuration at a spacing of 5.3~mm. Within each optical element, ortho-mode transducers absorb orthogonal incident polarization. The signal is split into the 90 and 150~GHz spectral bands by on-chip filters and then fed to one of the four detectors. 

Behind the detector stack lies the cold multiplexer~\cite{mccarrick_2021}, which is identical for MF and UHF. Briefly, microwave SQUID multiplexing ($\mu$mux)~\cite{Irwin04,Mates08,dober_2020} reads out the detector arrays at multiplexing factor of 910. The compact multiplexer, called the universal $\mu$mux module (UMM), was developed to occupy the same footprint as the detector array, and protrude minimally in the z-direction (i.e. away from the sky), for mechanical constraints at the telescope focal plane. Two pairs of RF chains and a single DC cable are used for each UFM. An outer shield, behind the cold multiplexer, is attached at the horn array flange, enclosing the inner UFM components which provides magnetic shielding~\cite{huber_2021}, mechanical protection, and minimizes stray light pickup. 

\begin{deluxetable}{cccccc}
\tablenum{1}
\tablecaption{Mid-frequency array target values for the Simons Observatory. The efficiency $\eta$ is at the horn aperture through detector absorption. The target NET is calculated using BoloCALC~\cite{hill2018} for a SAT array. Both 'baseline' and 'goal' sensitivity values are included following SO conventions~\cite{so_forecast_2019}. \label{tab:ufm}}
\tablewidth{0pt}
\tablehead{{Parameter} & {Definition} & {90~GHz} & {150~GHz} & {Units} & {Notes}}
\startdata
$P_\mathrm{sat}$ & saturation power     & 2.0--3.3  &  5.4--9.0   &    pW                 & performance requirement range \\ 
$\tau$           & time constant        &  0.4--1.1 &   0.3-1.0  &    ms                 & target range  \\
$R_\mathrm{n}$   & operating resistance &  2.6--7.2   &    2.6--7.2   &    m$\mathrm{\Omega}$ & target range \\ 
$T_\mathrm{c}$   & critical temperature & 160  &    160 &    mK                 & target      \\  
$G$              & thermal conductance  & 73  &   202  &    pW/K                & target  \\ 
$\eta$           & optical efficiency   &  70  &   70   &    $\%$               & target\\ 
$T_\mathrm{b}$   & bath temperature     & 100  &    100 &    mK                 & instrument target \\  
$\mathrm{NET}_\mathrm{CMB}$& array sensitivity &  3.4     & 4.3   &    $\mu$K$\sqrt{\mathrm{s}}$ & instrument performance target, baseline \\
$\mathrm{NET}_\mathrm{CMB}$& array sensitivity &  2.3     & 2.6   & $\mu$K$\sqrt{\mathrm{s}}$ & instrument performance target, goal \\
\enddata
\label{table_design}
\end{deluxetable}

\subsection{Specifications and requirements}

We discuss the targeted UFM performance including the specifics of the MF detector design parameters. The readout requirements and overall noise have been satisfied as previously described~\cite{mccarrick_2021}. 

The TES detector parameters are summarized in Table~\ref{table_design}. The TES critical temperature $T_\mathrm{c}$ is 160~mK. The UFMs are designed for a bath temperature $T_\mathrm{b}$ of 100~mK. The detector saturation powers $P_\mathrm{sat}$ are tuned for the expected loading, resulting in 2.6~pW (7.2~pW) for 90~GHz (150~GHz). The resulting targeted thermal conductance $G$ is 73~pW/K (202~pW/K), with an assumed fixed index $n$ of 3.7. The normal TES resistance is 8~m$\mathrm{\Omega}$ for both detector types. The time constants are tuned to be within an acceptable range for the beam and scan speed for both receivers to to 0.4-1.1~ms (0.3-1.0~ms). 

The strategy for MF array fabrication and acceptance is as follows: there is inherent variability in the fabrication of the devices and the design targets have evolved with increased testing and feedback. Thus, while the targets remain in place, a range of parameters are accepted so long as functionality and overall array sensitivity meet the necessary levels as set forth by the forecasting paper~\cite{so_forecast_2019}. To this end, we use as metrics the projected array noise-equivalent temperature (NET) and optical efficiency, which are somewhat degenerate with one another. The required MF array NET is 3.4~$\mu$K$\sqrt{\mathrm{s}}$ (4.3~$\mu$K$\sqrt{\mathrm{s}}$) in the SATs, and slightly higher for the LAT. The projected array NET, used as part of the acceptance criteria, is calculated using BoloCALC~\cite{hill2018}, incorporating measured detector parameters. 

\section{Results}
\begin{figure}[t!]
\begin{center}
\includegraphics[width=.9\linewidth]{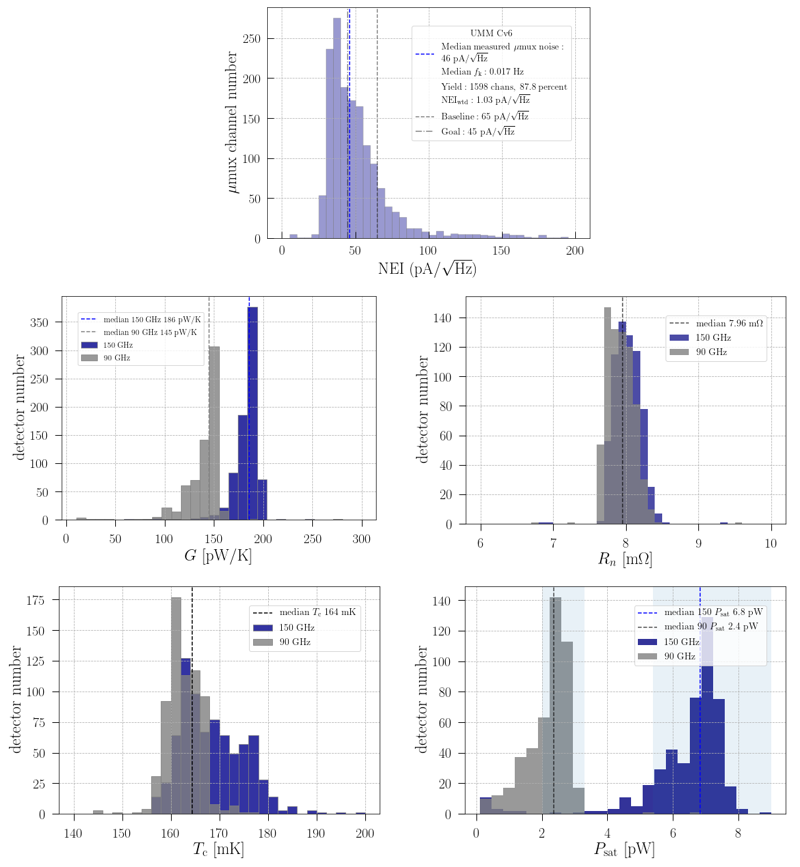}
\caption{({\it Top}) Readout noise from the multiplexer alone in units of NEI. The plots of detector parameters from mid-frequency UFMs. Due to the inherent variability introduced in fabrication, a range around the target design parameters are accepted. However, we require the projected on-sky array sensitivity to be better than the targeted baseline NET. ({\it Bottom}) Plots of measured detector parameters:  $G$ ({\it top left}), $R_\mathrm{n}$ ({\it top right}), $T_\mathrm{c}$ ({\it bottom left}), and $P_\mathrm{sat}$ ({\it bottom right}). Shadings in the measured $P_\mathrm{sat}$ plot indicate the targeted ranges. (Color figure online.)}
\end{center}
\label{fig_performance}
\end{figure}

We discuss measurements of the first MF UFMs for SO, to demonstrate typical MF detector performance. Ultimately, 26 MF UFMs will be fabricated, characterized, and deployed among the LATR and SATs. 

We briefly outline the testing setup here. For further details on the systematic testing program, see the concurrent publication~\cite{wang_2021}. The UFMs are cooled to 100~mK in laboratory dilution refrigerator (DR) cryostats, outfitted with dual-stage low-noise amplifiers on the output side of the RF chain. The SLAC Microresonator Radio Frequency (SMuRF)~\cite{henderson2017} is used as the corresponding warm electronics for the UFM; it is specifically designed to work with the cold $\mu$mux readout, providing both the RF signals ($\mu$mux probe tones) and DC signals for the readout (flux ramp) and TES bolometers (bias). 

The readout now routinely meets or exceeds the SO requirements in terms of both noise and yield. The cold multiplexer is generally measured alone before being coupled to an MF detector array. We show an example of the noise-equivalent current (NEI) from the multiplexer at the full 2x910 multiplexing factor in Fig.~\ref{fig_performance}. The median here is 46~$\mathrm{pA/\sqrt{Hz}}$, better than the SO requirement of 65~$\mathrm{pA/\sqrt{Hz}}$. 

We measure the detector parameters in the MF UFMs. A suite of standard tests extract the common TES bolometer parameters that define their sensitivity and noise. Example parameters from early MF UFMs are shown in Fig.~\ref{fig_performance}. The measured median $T_\mathrm{c}$ is 164~mK. The median saturation power $P_\mathrm{sat}$ is 2.4~pW (6.8~pW) at 90~GHz (150~GHz). The thermal conductance $G$, calculated with the also-fit index $n$, is approximately 145~pW/K (186 pW/K). We also plot the TES resistance $R_\mathrm{n}$, demonstrating the uniformity of the array. As described above, we project the on-sky NET using the measured parameters from the UFMs. The projected NET for a fairly typical, example MF UFM in the SAT is 3.3~$\mu$K$\sqrt{\mathrm{s}}$ (3.2~$\mu$K$\sqrt{\mathrm{s}}$), better than the baseline requirements and leaving margin to achieve the desired mapping speed.

\section{Discussion}

The UFMs for SO are agnostic with respect to integration into the SO small or large telescope receivers. All spectral bands share commonalities, with the MF and UHF being particularly similar. In this proceedings we have described the mid-frequency UFMs for SO, laid out the target parameters, discussed acceptance criteria, and presented in-laboratory measurements from initial MF UFMs. Ultimately, 26 mid-frequency UFMs will be deployed across the LAT and SATs, with first light expected in 2022. 

\begin{acknowledgements}
This work was supported in part by a grant from the Simons Foundation (Award $\#$457687, B.K.).
S.K.C. acknowledges support from NSF award AST-2001866. 
Z.B.H. is supported by a NASA Space Technology Graduate Research Opportunities Award.
Z.X. is supported by the Gordon and Betty Moore Foundation through grant GBMF5215 to the Massachusetts Institute of Technology.
Datasets for this article are available from the corresponding author on reasonable request and in accordance with the SO data policy. 
\end{acknowledgements}


\end{document}